\documentstyle[11pt]{article}\textheight 230mm\textwidth 150mm
            \newcommand{\be}{\begin{eqnarray}}
            \newcommand{\ee}{\end{eqnarray}}
            \newcommand{\eel}[1]{\label{#1}\end{eqnarray}}
\newcommand{\e}[1]{\label{e:#1}\end{eqnarray}}
                        \newcommand{\f}{f^{(\alpha)}}
     \newcommand{\eg}{{\em e.g.\ }}
            \newcommand{\ie}{{\em i.e.\ }}

            \newcommand{\del}{{\delta}}

           \newcommand{\ra}{{\rightarrow}}

            \newcommand{\beq}{\begin{quote}}
            \newcommand{\eq}{\end{quote}}
            
            \newcommand{\al}{\alpha}
            \newcommand{\ben}{\begin{enumerate}}
            \newcommand{\een}{\end{enumerate}}
            \newcommand{\bit}{\begin{itemize}}
            \newcommand{\ei}{\end{itemize}}
        \newcommand{\nn}{\nonumber}
            \newcommand{\r}[1]{(\ref{e:#1})}
            \newcommand{\edfl}[1]{\label{#1}\end{df}}
 \newcommand{\vb}{{\cal h}}
\newcommand{\hb}{{\cal i}}

\newcommand{\half}{\frac{1}{2}}
         
\begin{document} \begin{titlepage}
\noindent
G\"{o}teborg ITP 98-2, DAMTP-98-21, HKS-NT-FR-98/2-SE\\
31 March 1998\\
(Revised 22 March 1999)\\
\vspace*{5 mm}
\vspace*{10mm}
\begin{center}{\LARGE\bf A  general BRST approach to string theories\\ with zeta
function regularizations}\end{center}
\vspace*{3 mm}
\begin{center}
\vspace*{3 mm}

\begin{center}

Stephen Hwang\footnote{E-mail:
Stephen.Hwang@hks.se}, \\  \vspace*{2 mm} {\sl Karlstad University\\
S-65188 Karlstad, Sweden}
\\ \vspace*{7 mm}

Robert
Marnelius\footnote{E-mail: tferm@fy.chalmers.se},  \\ \vspace*{2 mm} {\sl
ITP, Chalmers
University of
Technology\\ G\"{o}teborg University\\ S-412 96  G\"{o}teborg, Sweden}\\
\vspace*{7 mm}  Panagiotis
Saltsidis\footnote{E-mail: P.Saltsidis@damtp.cam.ac.uk} \\
\vspace*{2 mm} {\sl
DAMTP, University of Cambridge\\ Silver St, Cambridge CB3 9EW, UK} \end{center}
\vspace*{18 mm}
\begin{abstract}
We propose a new general BRST approach to string and string-like theories which
have a wider range of applicability than \eg
 the conventional conformal field theory method. The method involves a
simple general
regularization of all basic commutators which makes all divergent sums to be
expressible in terms of zeta functions from which finite values then may be
extracted
in a rigorous manner. The method is particular useful in order to investigate
possible state space representations to a given model.
  The method is
applied to
three string models:
 The ordinary bosonic string, the
tensionless string and the conformal tensionless string. We also
investigate different
state spaces for these models. The tensionless string models are treated in
details. Although we mostly rederive known results they appear in a new fashion
which deepens our understanding of these models.  Furthermore, we believe
that our
treatment is more rigorous than most of the previous ones. In the  case of the
conformal tensionless string we  find a new solution for
$d=4$.
\end{abstract}\end{center}\end{titlepage}
\setcounter{page}{1}
\setcounter{equation}{0}
\section{Introduction and presentation of the method.}
The usual  operator formulation of  BRST quantization of string theories
are based on
the following ingredients: First specify
 a BRST invariant vacuum state and then normal
order the BRST operator in order to finally check in which dimension the
BRST operator is
nilpotent \cite{Ste}. Even the conformal field theory method is based on these
ingredients \cite{FMS}. These methods have been very successful when applied to
conventional string theories. However, there are models to which these
methods are
not applicable.  We have \eg the tensionless
string models
which do not have a conventional vacuum state
\cite{Is,GLS}. In this  paper we  present a new rigorous and very general
approach to the operator version
of BRST quantization, which not only makes it  possible to consistently
treat the conventional models but also  the
tensionless  string models. The
method is particularly useful in order to investigate possible state space
representations to a given model. The method seems, furthermore, to  cast
new light on
the BRST method as a whole.

What we   advocate here is the
following general procedure in the BRST quantization:
\begin{enumerate}
\item Construct a hermitian and formally nilpotent BRST operator $Q$.
\item Find a state space such that the properties above are true as
operator equations
in this space and for which the equation $Q|\phi\hb=0$ has non-trivial
solutions.
\item Use a general precise regularization of the basic commutators in
the above analysis.
\end{enumerate}
The idea to start with a nilpotent BRST operator and then look for possible
solutions
is of course very natural. However, usually one prescribes the setting first,
which is what
mathematicians would tell us to do. Usually the state space governs how we
construct the BRST charge.
But here
this is not required. We only require a formally nilpotent BRST charge to
start with.
Then we look for  possible state spaces. This freedom seems quite large.
However, it
is what a physicist likes to play with since physical models should also
allow for
our physical
intuition to act. The structure of possible solutions of such a general
prescription for finite degrees of freedom
was given in \cite{Princ} and further developed in \cite{Aux}. One may note
that a
hermitian, nilpotent BRST charge may be constructed for a very large class
of models
in terms of a power expansion in $\hbar$ and in the ghost fields \cite{BAF}. A
solution is naturally obtained in a Weyl ordered form.   However, these
properties
are only  formal, since we must also find a state space in which the BRST
charge makes
sense.  In order to do this
 in the case of infinite degrees of
freedom we must make use of a regularization procedure. For instance, in
the string
models   we encounter divergent sums. Such sums must be
regularized in some way. Here we consider a simple general regularization
of the
basic commutators, a regularization which will make all infinite sums to be
expressible as zeta functions. This regularization, which is presented in
section 2,
is therefore such that when the regulator is removed it will give finite results
through analytic continuation in all cases considered. This
makes it possible to rigorously compute all operator equations. Furthermore, and
which is important, it allows us  to investigate possible state spaces both
rigorously and efficiently. We emphasize that the  zeta function regularization
here is used in a much more general and precise form than what one usually
finds since
it here appears from one single regularization of the basic commutators.  (For
applications of zeta regularization and for literature on the subject, see \eg
\cite{Eliz}.)

Instead of giving a detailed prescription how the method is supposed to be
applied,
we treat three string models in details: the closed ordinary bosonic string, the
closed tensionless string and the closed conformal tensionless string. The
closed
ordinary bosonic string is mainly treated for pedagogical reasons. However,
it should
be interesting to see how we investigate alternative state space
representations. Our
main interest is in the tensionless strings. Here we give a rather exhaustive
analysis from which we are able to give precise results for all proposed
versions.

Critical dimensions in string theories appear in connection with a
particular state
space. In our method they appear through an inconsistency: The formally
nilpotent
BRST charge is {\em not} nilpotent or is not appropriate on the considered state
space. In our investigation of three string models we first look for a BRST
invariant
vacuum state. Then we check whether or not the formally nilpotent BRST charge is
nilpotent on this vacuum state. The conventional
closed bosonic string turns  out to be the most intricate example here.
The conventional vacuum state is investigated in section 3. The appropriate BRST
charge turns out to be the formally nilpotent charge shifted by a regulator
dependent
term (eq.\r{321}). This charge is, however,  only nilpotent at
$d=26$.  Thus, the standard results are obtained but in  a different  way than
usually. This treatment demonstrates how our method works and maybe it also
deepens
our understanding of this
well-known model. The second model we treat is the
tensionless string \cite{tension,dtwo,GRR}. In section 4
 we recover the known result of no critical dimension in
\cite{dtwo} for
one vacuum state, and in section 5 we find the critical dimension 26 as was
found in
\cite{GRR} for another
vacuum state.  Our final example is
the conformal tensionless string.  In section 6 we first investigate the
state space
considered in \cite{GLS}. Here we find that the BRST charge is nilpotent in any
dimension. However, the vacuum state is only BRST
invariant in two dimensions and we have not found any BRST
invariant states in other
dimensions. In section 7 we also consider another state space, which to our
knowledge has not been
considered previously. In this state space the BRST charge is again
nilpotent in
arbitrary dimensions, but the vacuum state is only BRST invariant in $d=4$.
For other values of
$d$ we have not found any BRST invariant states in this state space. In
section 7 we
investigate  also the alternative vacuum state considered
in \cite{Is} in which case we find that the BRST charge is not nilpotent in any
dimensions and, hence, that there is no consistent BRST treatment at all,
which is in
agreement with the result of \cite{Is}. In  sections 8 and 9
we investigate alternative vacua also for the ordinary bosonic string. Note
that
unlike some previous calculations  our regularization makes all
results finite which makes us believe that the treatment given here is
in general more
rigorous than previous ones.

Normally a consistent BRST quantization requires   BRST
invariant
states with positive norms. In fact, such a condition should really be
inserted into
the condition 2 above. However, this would then exclude all models we treat
except
possibly the ordinary bosonic string. The reason is that
all our calculations are performed in the minimal sector which
is the sector with no dynamical
Lagrange multipliers and antighosts. An artefact of this sector is
that one cannot in general work
on a truly inner product space. Consistent BRST invariant states
$|\phi\hb$ requires here the existence
of a dual BRST invariant state $|\bar{\phi}\hb$ satisfying the condition
\cite{Princ}
\be
&&\vb\bar{\phi}|\phi\hb \mbox{\rm  is finite and} \neq 0.
\e{1}
This condition restricts the possible solutions. To work on  truly inner product
 spaces
one has to consider a BRST quantization in the complete sector
with dynamical Lagrange multipliers and antighosts \cite{Gauge}. However,
inconsistent solutions in the minimal sector will remain inconsistent in
the complete
sector. Furthermore, solutions containing BRST invariant negative norm
states in the
minimal sector will retain these in the complete sector. In section 9 we
present a
state space representation for the conventional bosonic
string model which yields a finite set of BRST invariant states, which however
are shown to contain negative norm states. The same is true for the special
solutions of the tensionless string in section 5. Both these options have
therefore to be excluded.   The
consistency of the tensionless string model in
section 4 seems  unclear, since it has a vacuum solution which is not
associated with
any oscillators. The BRST invariant states are not inner product states at
all in the minimal sector. Although we expect there exist a positive normed
solution
in the complete sector, this remains to be investigated.

It is of course also possible to investigate the
fermionic extensions of the models considered here
\cite{GRR,Pan}. We expect such calculations to be quite straightforward
using the
method we present here. One may also consider other theories
like \eg brane-theories \cite{Pan2} and conventional field theories. We
believe
 that also
here our
method should be important for rigorous results.

\setcounter{equation}{0}
\section{Our regularization}
In string theory the string coordinates $X^{\mu}(\tau,\sigma)$ and the
 corresponding conjugate momenta $P_{ \mu}(\tau,\sigma)$ satisfy the basic
nonzero
equal-time commutator (we suppress $\tau$ in the following)
\be
&&[X^{\mu}(\sigma), P_{\nu}(\sigma')]=i\del^{\mu}_{\nu}\del(\sigma-\sigma').
 \e{121}
Since we shall only consider closed strings we let $X^{\mu}(\sigma)$ and
$P_{\mu}(\sigma)$ be periodic functions with period $\pi$. Thus, we may set
($\sum_n\equiv\sum_{n=-\infty}^{\infty}$ in the following)
\be
&&X^{\mu}(\sigma)={1\over
\sqrt{\pi}}\sum_{n}x_n^{\mu}e^{-2in\sigma},\quad
P^{\mu}(\sigma)=
{1\over \sqrt{\pi}}\sum_{n}p_n^{\mu}e^{-2in\sigma}
 \e{122} and
replace \r{121} by
\be
&&[x^{\mu}_n, p_{m\nu}]=i\del^{\mu}_{\nu}\del^0_{n+m}.
\e{123}
The delta function in \r{121} is then the periodic delta function
\be
&&\del(\sigma-\sigma')={1\over
\pi}\sum_{n}e^{2in(\sigma-\sigma')}.
\e{124} It is the appearance
of a delta function in the commutator \r{121} that causes infinities in the
quantum
string theory. In order to handle these infinities in a well defined manner
we have
to regularize the delta function, \ie we have to make it an ordinary well
defined
function. We have  therefore to consider a modified or regularized
commutator. We
choose it to be of the form
\be
&&[x^{\mu}_n, p_{m\nu}](s)=i\del^{\mu}_{\nu}\del^0_{n+m} f(|n|,s),
\e{125} where
$f(|n|,s)$ is a  real function which satisfies the condition
\be
&&f(|n|,0)=1.
\e{126}
It can only depend on the absolute value of $n$ in order for
$x_n$ and $p_n$ to retain
their hermiticity properties from \r{122} even in the regularized case, \ie
$(x_n)^\dag=x_{-n}$ and
$(p_n)^\dag=p_{-n}$. The choice of regulator function $f(|n|,s)$ is
dictated by two
considerations. Firstly, in the computations that will follow \eg a
calculation of the
BRST nilpotency, the regulator should lead to a finite result as the
regulator is
removed ($s\rightarrow 0$).  Secondly, we would like our regulator
to be as general as possible.
A precise choice of $f(|n|,s)$ which satisfies these criteria is
\be
&&\f(|n|,s)=\left\{\begin{array}{ll}
(|n|+\al)^{-s},&|n|\geq-A+1\\
1,& |n|\leq-A
\end{array}\right.,
\e{127}
where $\al$ is a real constant, which may be chosen to have any value, and
where $A$
is the closest integer to $\al$ satisfying
$A\geq\al$. This choice will make all infinite sums to be expressible in
terms of
zeta functions (see below). This in turn will allow us to rigorously treat all
infinite sums. Other regularizations are of course possible to use and
should yield
equivalent results. However, we think that the choice \r{127} is the most
general one.

In the following BRST treatment of the
various string models we have also the fermionic
ghost variable
$c^I(\sigma)$ and
$b^I(\sigma)$. Their basic commutators must be regularized in exactly the
same way as
in \r{125}. For ghosts $c_n^{I},b_n^{I}$ (where $I$ label
different types of ghosts)  we have :
\be
&&[b_m^{I},c_n^{J}](s)=\delta^0_{m+n}\delta^{IJ}\f(|n|,s).
\e{1271}
The same is true for all other canonical variables one introduces. (We use
graded
commutators throughout.)

When calculating commutators we often get infinite sums of the form
$\sum_n \f(|n|,s)$
which converge for $s>1$. However, by analytic continuations they may yield
a finite
value for $s<1$ and in particular for $s=0$. The choice \r{127} makes these
sums
expressible in terms of zeta
functions. We have
\be
&&\sum_{n} \f(|n|,s)=\left\{\begin{array}{ll} 1-2A+
2\zeta(s|1+\al-A),&\al\leq0\\
-\al^{-s}-2\sum_{n=0}^{A-2}(n+1+\al-A)^{-s}+2\zeta(s|1+\al-A),& \al>0
 \end{array}\right.\nn\\
\e{128}
where $\zeta(s|a)$ is the Hurwitz zeta function defined for $0<a\leq 1,\ s>1$ by
\cite{WW} \be &&\zeta(s|a)\equiv\sum_{n=0}^\infty{1\over (n+a)^s}.
 \e{129}
By analytic continuation one finds the following value at
 $s=0$ ($0<a\leq 1$)  \cite{WW}
 \be
 &&\zeta(0|a)=\half-a,
\e{1291}
which implies that (notice that the finite sums for $\al\leq0$ and $\al>0$
are equal
for $s=0$ in \r{128})
\be
&&\sum_{n} \f(|n|,0)=-2\al.
\e{1292}
We have similarly \eg (we set $s=0$ for the finite sums)
\be
\sum_{n=1}^\infty
n\f(|n|,s)=&&\zeta(s-1|1+\al-A)-\al\zeta(s|1+\al-A)+\half
A(A-1),\nn\\
\sum_{n=1}^\infty
n^2\f(|n|,s)=&&\zeta(s-2|1+\al-A)-2\al\zeta(s-1|1+\al-A)+\al^2\zeta(s|1+\al-A)
-\nn\\
&&-{1\over6}A(A-1)(2A-1),
\e{1293}
which implies
\be
&&\lim_{s\ra0}\sum_{n=1}^\infty
n\f(|n|,s)=\half\al^2-{1\over
12},\qquad
\lim_{s\ra0}\sum_{n=1}^\infty
n^2\f(|n|,s)=-{1\over3}\al^3,
\e{1294}
since $\zeta(-1|a)=(6a-6a^2-1)/12$ and $\zeta(-2|a)=a(3a-2a^2-1)/6$.

The
regularization \r{125} and \r{127} correspond by \r{121} to the following
regularized delta
function
\be
&&
\del_s(\sigma-\sigma')={1\over
\pi}\left\{\sum_{n}\f(|n|,s)e^{2in(\sigma-\sigma')}\right\},
\e{1295}
which is a well defined function for $s>1$ and by eq. \r{1292} gives a
regularization
such that $\lim_{s\rightarrow 0}\delta_s(0)={-2\alpha\over \pi}$. Note here that
depending on the value of $\alpha$ we can get any value in this limit. This
is one
argument supporting our belief that our regularization gives the most
general result
possible. Notice also that $\lim_{s\rightarrow 0}\delta^{(k)}_s(0)$ is
finite for
any order of derivative $k$.\\
\setcounter{equation}{0}
\section{The bosonic string}
As an illustration of our method we first treat the ordinary bosonic string.
 The bosonic string is classically characterized by the constraints
\be
&&{1\over 4T}(P+TX')^2(\sigma) =0,\quad {1\over 4T}(P-TX')^2(\sigma)=0,
\e{301}
where $T$ is the string tension.
The Fourier modes of the corresponding hermitian constraint operators are
\be &&L_n \equiv \half \sum_{k} \alpha_{n-k}\cdot
\alpha_k,\qquad K_n \equiv
\half
\sum_{k}
\tilde{\alpha}_{n-k}\cdot \tilde{\alpha}_k,
\e{302}
where
\be
&&\alpha_{n}^{\mu}\equiv
\left
(\frac{1}{2\sqrt{T}}p_n^{\mu}-i\sqrt{T}nx_{n}^{\mu}\right ), \nonumber\\
&&\tilde{\alpha}_{n}^{\mu} \equiv  \left (\frac{1}{2\sqrt{T}} p_{-n}^{\mu}-
i\sqrt
{T}n x_{-n}^{\mu}\right ),
\e{303}
in terms of the $x_n$ and $p_n$ modes in \r{122}. The regularized
commutator \r{125}
with the regularized function \r{127} imply
\be
&&[\al^{\mu}_n, \al^{\nu}_m](s)=n\eta^{\mu\nu}\f(|n|,s)\del^0_{n+m},\qquad
[\al^{\mu}_n,
\tilde{\al}^{\nu}_m](s)=0,\nn\\ &&[\tilde{\al}^{\mu}_n,
\tilde{\al}^{\nu}_m](s)=n\eta^{\mu\nu}\f(|n|,s)\del^0_{n+m},
\e{304}
where $\eta^{\mu\nu}$ is a  Minkowski metric,
$\mbox{diag}(\eta^{\mu\nu})=(-1,+1,+1,+1)$. We consider now a corresponding
BRST theory in which  the manifestly hermitian
BRST charge is given by
\be {{Q}}&=& \sum_{k}  (
L_{-k}{c}_{k}^{L}+K_{-k}c_{k}^{K})+\nonumber\\  &&+ {1\over
4}\sum_{k,l}(k-l)\left(
{b}_{k+l}^{L}{c}_{-l}^{L}{c}_{-k}^{L}+b_{k+l}^{K}c_{-l}^{K}c_{-k}^{K}-
{c}_{-k}^{L}{c}_{-l}^{L}{b}_{k+l}^{L}-   c_{-k}^{K}c_{-l}^{K}b_{k+l}^{K}\right),
\e{305}
where $c^L_k$, $b^L_k$ and $c^K_k$, $b^K_k$ are fermionic ghost modes satisfying
\r{1271}. In the
$s\ra 0$ limit this BRST charge is formally nilpotent. A consequence of this
nilpotency  is that the extended constraints, $[Q, b^I_k]$, satisfy a closed
algebra
without any central extensions. The extended constraints to $L_n$ and $K_n$ are
\be
&&\tilde{L}_n \equiv\left [Q,b^{L}_n \right ]=L_n -
\half\sum_{k}(k+n)(
c^{L}_{k}b^{L}_{n-k}-b^{L}_{n-k}c^{L}_{k}),\\
&&\tilde{K}_n
\equiv
\left [Q,b^{K}_n \right ]=
K_n -
\half \sum_{k}(k+n)
(c^{K}_{k}b^{K}_{n-k}-b^{K}_{n-k}c^{K}_{k}).
\e{306}
For non-zero $s$ we have \eg the following crucial commutator
\be
&&\left [\tilde{L}_{m},\tilde{L}_{-m}\right ](s)=\nonumber\\&&
\half m\sum_{k}\left(\alpha_{-k}\cdot \alpha_k+k
(c^{L}_{-k}b^{L}_{k}-b^{L}_{k}c^{L}_{-k})\right)\left(f^{(\al)}(|m+k|,s)+
f^{(\al)}(|m-k|,s)\right)+\nonumber\\
&&+\half\sum_{k}\left( k\alpha_{-k}\cdot \alpha_k+
(2m^2-k^2)(b^{L}_{k}c^{L}_{-k}-c^{L}_{-k}b^{L}_{k})\right)\left(f^{(\al)}(|m
+k|,s)-
f^{(\al)}(|m-k|,s)\right),\nonumber\\&&
\e{307}
which in the limit $s\ra 0$ becomes
\be
&&\left [\tilde{L}_{m},\tilde{L}_{-m}\right ](0)=2m\tilde{L}_{0},
\e{308}
which is consistent with the fact that the BRST charge \r{305} is nilpotent
in the
$s\ra 0$ limit. Now the crucial point is that the $s\ra 0$ limit has no meaning
before we specify on which state space the operators act. Below we
show that
the conventional choice of state space imply the expected result that we have a
nilpotent BRST charge only in spacetime dimensions
$d=26$ in the $s\ra 0$ limit.

The standard choice of a vacuum state, $|0\hb$,
satisfies
\be
&&\al_m|0\hb=\tilde{\al}_m|0\hb=0,\qquad \forall m>0.
\e{309}
In order for this vacuum to be BRST invariant, $Q|0\hb=0$, we must require the
consistency conditions
\be
[Q,\alpha^{\mu}_m] |0\rangle = 0&&\Rightarrow
m\f(|m|,s)\sum_{k=m}^{\infty}c^{L}_{k}\alpha^{\mu}_{m-k}|0\rangle =0\nonumber\\
&&\Rightarrow  c^{L}_{m}|0\rangle
=0,\qquad \forall m>0,\nonumber\\
\left [{{Q}},\tilde{\alpha}^{\mu}_m\right ]|0\rangle =0&&\Rightarrow
 m\f(|m|,s)\sum_{k=m}^{\infty}c^{K}_{k}\tilde{\alpha}^{\mu}_{m+k}|0\rangle
=0\nonumber\\ &&\Rightarrow {c}^{K}_{m}|0\rangle =0,\qquad \forall m>0.
\e{310}
These  conditions in turn allow for the additional conditions
\be {b}^{L}_m|0\rangle=b^{K}_m|0\rangle=0, \qquad\forall m\geq 0,
\e{311}
for which the consistency conditions are
\be
\left [{{Q}},b^{L}_m \right ] |0\rangle &\equiv& \tilde{L}_m|0\rangle=0,
\qquad
\forall m\geq 0,\nonumber\\
\left [{{Q}},b^{K}_m \right ] |0\rangle &\equiv& \tilde{K}_m|0\rangle=0,
\qquad
\forall m\geq 0.
\e{312}
They are satisfied for $m>0$, but for $m=0$ the situation is unclear.
For non-zero $s$ we have
\be
&&\tilde{L}_0(s)|0\hb=\half\left(\al_0^2+(d-2)\sum_{k=1}^\infty
k\f(\mid k\mid ,s)\right)|0\hb,
\e{313}
which in the $s\ra 0$ limit leads to the following finite expression
\be
&&\tilde{L}_0|0\hb=\half\left(\al_0^2+(d-2)(\half\al^2-{1\over
12})\right)|0\hb,
\e{314}
where we have made use of the relation \r{1294}.
 The conditions \r{312} with the property \r{314} and a
similar one for $\tilde{K}_n$ require the vacuum state to be an eigenstate
to the
momentum operator $p_0^{\mu}$ with an eigenvalue which depends on the
parameter $\al$
in the regularization function \r{127}. (Notice that \r{303} implies
$\al^2_0=\tilde{\al}^2_0=p_0^2/4T$.) This is an unsatisfactory result. It
means that
the conventional vacuum state is not BRST invariant under the formally
nilpotent BRST
charge above.  In fact, $Q$  is not even  nilpotent on the conventional
vacuum state.

That the BRST operator \r{305} is {\em not} nilpotent on the above vacuum
state may be
seen by calculating
 the commutator $[\tilde{L}_{m},\tilde{L}_{-m}]$ on the
vacuum state for non-zero $s$. We find from eq. \r{307}
\be
&\left[\tilde{L}_{m},\tilde{L}_{-m}\right](s)|0\hb=&
\left(m\f(|m|,s)\alpha^{2}_0
+\half d\sum_{k=1}^{m}k(m-k)
\f(|k|,s)\f(|k-m|,s)-\right.\nn\\ &&-
\left.\sum_{k=1}^{m}(2m-k)(k+m)\f(|k|,s)\f(|k-m|,s)\right)|0\hb.
\e{315}
In the $s\ra 0$ limit  the right-hand side becomes
\be
&&\left(m\al_0^2+{1\over12}(d-26)m^3-{1\over12}(d-2)m\right)|0\hb,
\e{316}
which only is zero if $d=26$ and if the eigenvalue of $\al_0^2$ is 2, which is a
regulator independent condition. The reason for the
different results from \r{308} and \r{314} is due to the fact that on the
right-hand
side of \r{307} the factor $\f(|m+k|,s)-\f(|m-k|,s)$, which is zero in
the $s\ra0$ limit, is multiplied by an operator which is infinite on the above
vacuum state. This means that  $\tilde{L}_0$
and $\tilde{K}_0$ are not zero on the vacuum state $|0\hb$ when \r{316} is
zero.  The  solution of this dilemma is found when we rewrite the
right-hand side
of the
commutator
\r{307} by means of the regularized commutators \r{304} as follows
\be
&&\left [\tilde{L}_{m},\tilde{L}_{-m}\right ](s) =\nn\\
&&m\sum_{k=1}^{\infty}\left
(\alpha_{-k}\cdot\alpha_{k} +kb^{L}_{-k}c^{L}_{k}+kc^{L}_{-k}b^{L}_{k} \right )
\left
(\f(|m+k|,s)+\f(|m-k|,s)
\right )+\nn\\ &&+
\sum_{k=1}^{\infty}\left
(k\alpha_{-k}\cdot\alpha_{k}+(k^2-2m^2)
(b^{L}_{-k}c^{L}_{k}+c^{L}_{-k}b^{L}_{k})
\right ) \left (\f(|m+k|,s)-\f(|m-k|,s)\right )+\nn\\ &&+m\f(|m|,s)\alpha^{2}_0
+\half d\sum_{k=1}^{m}k(m-k)
\f(|k|,s)\f(|k-m|,s)-\nn\\ &&-
\sum_{k=1}^{m}(2m-k)(k+m)\f(|k|,s)\f(|k-m|,s).
\e{317}
This expression implies
\be
&\left [\tilde{L}_{m},\tilde{L}_{-m}\right ] = &
2m\left (\half \alpha^{2}_0 +\sum_{k=1}^{\infty}(\alpha_{-k}\cdot\alpha_{k}
+kb^{L}_{-k}c^{L}_{k}+kc^{L}_{-k}b^{L}_{k})\right)+\nonumber\\
&&+\frac{1}{12}(d-26)m^{3}-\frac{1}{12}(d-2)m
\e{318}
in the $s\ra 0$ limit in agreement with \r{315} and \r{316}.  This may be
rewritten
 as
follows
\be
&&\left [\tilde{L}_{m},\tilde{L}_{-m}\right
]=2m\tilde{L}_{0}+{1\over12}(d-26)m^3-\half m(d-2)\al^2,
\e{319}
where
\be
&&\tilde{L}_{0}=\half \alpha^{2}_0
+\sum_{k=1}^{\infty}(\alpha_{-k}\cdot\alpha_{k}
+kb^{L}_{-k}c^{L}_{k}+kc^{L}_{-k}b^{L}_{k})+\half(d-2)\left(\half\al^2-
\frac{1}{12}\right)
\e{320}
is the original $\tilde{L}_{0}$ rewritten for finite $s$ and taking
the limit
$s\ra0$. Eq.\r{319} demonstrates the inconsistency with a nilpotent BRST charge
obtained above and the reason why the conventional vacuum state is not BRST
invariant. The remedy is obvious and expected: first we notice that we may get a
closed algebra at
$d=26$ if we  redefine the extended constraint
$\tilde{L}_{0}$. This in turn may be  accomplished by a
redefinition of the original BRST charge: Simply replace $Q$ in \r{305} by
\be
&&Q'=Q-{d-2\over 4}\al^2(c_0^L+c_0^K),
\e{321}
which is not formally nilpotent for  $d\neq2$. The corresponding extended
constraints are
\be
&&\tilde{L}'_{m}=[Q', b^L_m]=\left\{\begin{array}{l}\tilde{L}_{0}-{1\over
4}(d-2)\alpha^2,\;m=0\\ \tilde{L}_{m}, \;m\neq0\end{array}\right. .
\e{322}
Note that $\tilde{L}'_{0}$ in distinction to $\tilde{L}_{0}$ in \r{320} is
independent of the
regulator parameter
$\alpha$. We have
\be
&&\tilde{L}'_{0}=\half \alpha^{2}_0
+\sum_{k=1}^{\infty}(\alpha_{-k}\cdot\alpha_{k}
+kb^{L}_{-k}c^{L}_{k}+kc^{L}_{-k}b^{L}_{k})-{(d-2)\over 24},
\e{324}
 The spectrum is therefore regulator independent as it should be.
For $d=26$
 we find now from \r{319}
\be
&&\left [\tilde{L}'_{m},\tilde{L}'_{-m}\right
]=2m\tilde{L}'_{0},
\e{323}
which is consistent with a nilpotent $Q'$  for
$d=26$, and that the conventional BRST vacuum is BRST invariant under $Q'$.

In sections 8 and 9 we investigate other vacuum states for the bosonic
string theory.

\setcounter{equation}{0}
\section{The bosonic tensionless string} The bosonic tensionless
string (see \cite{tension}) is characterized by the constraints \be
&&P^{\mu}(\sigma)P_{\mu}(\sigma)=0,\quad P^{\mu}(\sigma)X^{'}_{\mu}(\sigma)=0.
\e{41}
These constraints follow from the bosonic string  by dropping the term
$T^2(X^{\prime}(\sigma))^2$ which is assumed to be negligible in the
 $T\,\ra\,0$ limit.
The Fourier modes of the corresponding hermitian constraint
operators are
\be
&&\phi^{-1}_{n}\equiv\half \sum_{k}p_{k} \cdot
p_{n-k}, \quad\phi^{L}_{n}\equiv-i\half
\sum_{k}k\left(x_{k}\cdot
p_{n-k} + p_{n-k}\cdot x_{k}\right).
\e{42}
A formally nilpotent BRST charge operator is here given by
\be
{{Q}}&=& \sum_{k}  (
\phi_{-k}^{-1}c_{k}^{-1}+\phi_{-k}^{L}c_{k}^{L})-
\half\sum_{k,l}(k-l)\left(c_{-k}^{-1}c_{-l}^{L}b_{k+l}^{-1}+\right. \nonumber\\
&&\left.+b_{k+l}^{-1}c_{-k}^{-1}c_{-l}^{L}+\half
c_{-k}^{L}c_{-l}^{L}b_{k+l}^{L}+\half
b_{k+l}^{L}c_{-k}^{L}c_{-l}^{L}\right).
\e{43}  We can  check the nilpotency of  ${ Q}$ by calculating the
algebra of
the extended constraints given by
 \be
&&\tilde{\phi}^{-1}_n\equiv[{{Q}},b_n^{-1}]=\phi^{-1}_{n}-
\sum_{k} (n+k)c^{L}_{k}b^{-1}_{n-k}\nn\\
&&\tilde{\phi}^L_n\equiv[{{Q}},b_n^L]=\phi^{L}_{n}-
\half\sum_{k}(k+n)\left(c^{-1}_{k}b^{-1}_{n-k}-b^{-1}_{n-k}c^{-1}_{k}
+c^{L}_{k}b^{L}_{n-k}-b^{L}_{n-k}c^{L}_{k}\right).
\e{44}
A straight-forward calculation of the commutators yields
\be
&&[\tilde{\phi}^{-1}_m,\tilde{\phi}^L_n]=(m-n)\tilde{\phi}^{-1}_{m+n},\qquad
[\tilde{\phi}^{L}_m,\tilde{\phi}^L_n]=(m-n)\tilde{\phi}^{L}_{m+n}.
\ee
Hence, we conclude that the
BRST charge \r{43} is nilpotent.

We now look  for a possible BRST invariant vacuum state.
Following \cite{GLS,dtwo} we consider a vacuum state defined
by
\be
&&p_n^\mu|0\hb=b_n^{-1}|0\hb=b_n^L|0\hb=0,\quad \forall n\;.
\e{441}
(The crucial part is the first conditions.  They may be viewed as the
$T\ra0$ limit
of \r{309} using \r{303}.) In order for this vacuum state to be BRST
invariant it has
to satisfy the consistency conditions
\be
&&[{{Q}},p_n^\mu]|0\hb=[{{Q}},b_n^{-1}]|0\hb=
 [{{Q}},b_n^L]|0\hb=0,\quad \forall n \;,
 \e{45}
 where
\be
&&[{{Q}},p^\mu_n]=-n\sum_{k} p_{n-k}^\mu c^L_{k}.
\e{46}
The first two conditions and the last one
 for $n\neq0$ are easily seen to be satisfied
due to \r{441}. The only nontrivial condition is the last one for $n=0$.
However, we have
\be
&&\tilde{\phi}^L_0|0\hb\equiv[{{Q}},b_0^L]|0\hb=\left(\phi^{L}_{0}-
\half\sum_{k}k[c^{-1}_{k}b^{-1}_{-k}-
b^{-1}_{-k}c^{-1}_{k}+c^{L}_{k}b^{L}_{-k}-b^{L}_{-k}c^{L}_{k}]\right)|0\hb=\nn\\
&&=-\half\sum_kkp\left(i[p_{-k}^{\mu},
x_{k\mu}]-[b^{-1}_{-k},c^{-1}_{k}]-[b^{L}_{-k},c^{L}_{k}]\right)|0\hb=
\nn\\ &&=-\half(d-2)\sum_kk\f(|k|,s)|0\hb=0.
\e{47}
(It is zero for any choice of regulator function $f(|k|,s)$!). It is then
easily seen that
\be &&{{Q}}|0\hb=0,
\e{48}
which means that $|0\hb$ defined by \r{441} is a BRST invariant vacuum state
for any
dimension $d$.
We must finally check that the BRST charge is nilpotent on the vacuum state.
This may be accomplished by checking the algebra of the extended
constraints. The
most non-trivial one is $[\tilde{\phi}^L_m,\tilde{\phi}^L_{-m}]$ for which
we find
\be
&&[\tilde{\phi}^L_m,\tilde{\phi}^L_{-m}](s)={i\over
2}\sum_k(k+m)kf^{(\al)}(|k+m|,s)(x_{-k}\cdot p_k-x_{k}\cdot p_{-k}+\nn\\
&&+ p_k\cdot x_{-k}-
p_{-k}\cdot x_k)+\half\sum_k(k+2m)(k-m)f^{(\al)}(|k+m|,s)(b_{-k}^{-1}c_{k}^{-1}
-b_{k}^{-1}c_{-k}^{-1}+\nn\\&&+c_{-k}^{-1}b_{k}^{-1}-c_{k}^{-1}b_{-k}^{-1}+
b_{-k}^{L}c_{k}^{L}
-b_{k}^{L}c_{-k}^{L}+c_{-k}^{L}b_{k}^{L}-c_{k}^{L}b_{-k}^{L}).
\e{49}
A straightforward calculation
yields
that
$[\tilde{\phi}^L_m,\tilde{\phi}^L_{-m}](0)|0\hb$ is zero, which is
consistent with eq.
\r{47}. To conclude, within our regularization scheme  we have shown that there
exists a BRST invariant vacuum state in any dimension. Furthermore, the BRST
charge is
nilpotent in the state space containing this vacuum state. This result is in
agreement with the results of ref.\cite{dtwo}. Note that there exists a
dual vacuum
state $|\bar{0}\hb$ satisfying $\vb\bar{0}|0\hb$ finite and different from zero.
$|\bar{0}\hb$ satisfies
$x_n^{\mu}|\bar{0}\hb=c_n^{-1}|\bar{0}\hb=c_n^L|\bar{0}\hb=0$
for all $n$ together with their consistency conditions.
\setcounter{equation}{0}
\section{Alternative quantization of the bosonic tensionless string}
Instead of a vacuum state satisfying \r{441} we follow \cite{GRR} and consider
\be
&&p_m^{\mu}|0\hb=x_m^{\mu}|0\hb=0,\quad m>0.
\e{401}
The consistency conditions are
\be
&&[Q, p_m^{\mu}]|0\hb=[Q, x_m^{\mu}]|0\hb=0,\quad m>0,
\e{402}
and they require
\be
&&c_m^{-1}|0\hb=c_m^{L}|0\hb=0,\quad m>0
\e{403}
for which
\be
&&[Q, c_m^{-1}]|0\hb=[Q, c_m^{L}]|0\hb=0,\quad m>0
\e{404}
are automatically satisfied. This vacuum is then ghost fixed by
\be
&&b_m^{-1}|0\hb=b_m^{L}|0\hb=0,\quad m\geq0.
\e{405}
The corresponding consistency conditions,
\be
&&\tilde{\phi}_m^{-1}|0\hb=\tilde{\phi}_m^{L}|0\hb=0,\quad m\geq0
\e{406}
are automatically satisfied for $m>0$. For $m=0$ we have
\be
&&\tilde{\phi}_0^{-1}|0\hb=0\quad\Leftrightarrow\quad p_0^2|0\hb=0.
\e{407}
For such a condition to be meaningful the vacuum state should be an
eigenstate to
$p_0^{\mu}$, \ie $p_0^{\mu}|p\hb=p^{\mu}|p\hb$. Eq.\r{407} requires then
the vacuum
to be massless ($p^2=0$). For $m=0$ we have also
\be
&&\tilde{\phi}_0^{L}|p\hb=(d-2)(\half\al^2-{1\over 12})|p\hb,
\e{408}
which  satisfies \r{406} only for $d=2$. For $s\neq0$ we have
\be
&&\tilde{\phi}_0^{L}(s)=i\sum_{k=1}^\infty k(x_{-k}\cdot p_{k}-p_{-k}\cdot
x_{k})+
\sum_{k=1}^\infty k(b^{-1}_{-k}\cdot c^{-1}_{k}+c^{-1}_{-k}\cdot
b^{-1}_{k}+b^{L}_{-k}\cdot c^{L}_{k}+\nn\\
&&+c^{L}_{-k}\cdot
b^{L}_{k})+(d-2)\sum_{k=1}^\infty kf^{(\al)}(|k|,s).
\e{409}
From the commutator \r{49} we find
\be
&&\lim_{s\ra0} [\tilde{\phi}_m^{L}, \tilde{\phi}_{-m}^{L}](s)|p\hb={1\over
6}\left((d-26)m^3-(d-2)m\right)|p\hb.
\e{410}
In fact, in the $s\ra0$ limit we have
\be
&&[\tilde{\phi}_m^{L},
\tilde{\phi}_{-m}^{L}]=2m\tilde{\phi}_0^{L}-\half\al^2(d-2)m+{1\over
6}(d-26)m^3,
\e{411}
where $\tilde{\phi}_0^{L}$ on the right-hand side is the $s\ra0$
limit of
\r{409}. If we redefine $Q$ by
\be
&&Q'\equiv Q-{1\over4}\al^2(d-2)c_0^L,
\e{412}
then the extended constraint operator $\tilde{\phi}_0^{L}$ is replaced by
\be
&&{\tilde{\phi}}_0^{'L}=\tilde{\phi}_0^{L}-{1\over 4}\al^2(d-2).
\e{413}
This operator together with $\tilde{\phi}_m^{L}$, $m\neq0$ satisfy then
 an anomaly free algebra for  $d=26$ from which we conclude that $Q'$ is
nilpotent for
$d=26$. However, from \r{408} we have
\be
&&{\tilde{\phi}}_0^{'L}|p\hb=-{1\over 12}(d-2)|p\hb.
\e{414}
It follows that we have no BRST invariant vacuum of the above form for $d=26$.
However, there are BRST invariant states in $d=26$. In \cite{GRR2} it was
shown that
there are massless states with spin 0,1 and 2. In section 9 we show that
these states
do not have positive norms.  (From the results of
\cite{GRR} it seems that a BRST invariant vacuum state only exists in the Ramond
sector of the spinning tensionless string in which case the critical
dimension is 10.)

\setcounter{equation}{0}
\section{The bosonic conformal string}
The conformal string is a tensionless string which is made manifestly
conformally invariant
\cite{Is, GLS}. By adding two extra dimensions, one timelike and one
spacelike, one
forms new coordinates that transforms as $SO(d,2)$ vectors. By means of
$SO(d,2)$
invariant constraints one obtains then an $SO(d,2)$ conformally invariant
formulation
by construction. Let $X^M=(X^\mu, X^+, X^-)$ be the new coordinate vector
where the
metric of the new coordinates is $\eta_{++}=\eta_{--}=0$,
$\eta_{+-}=\eta_{-+}=1$.
Classically the constraints are
\be
&&\Phi^{-1}(\sigma)\equiv P^{M}(\sigma)P_{M}(\sigma)=0,\quad
\Phi^{0}(\sigma)\equiv P^{M}(\sigma)X_{M}(\sigma)=0,\nonumber\\
&&\Phi^{1}(\sigma)\equiv X^{M}(\sigma)X_{M}(\sigma)=0,\quad
\Phi^{L}(\sigma)\equiv P^{M}(\sigma)X^{'}_{M}(\sigma)=0,
 \e{51}
 and they reduce to
the constraints \r{41} of the tensionless string by means of the gauge fixing
conditions
\be
P^{+}(\sigma)=0,\qquad X^{+}(\sigma)-1 =0.
\e{52}
(A corresponding construction for particles were given in \cite{MN}.) The
hermitian
BRST charge operator is given by
\be
{{Q}}&=& \sum_{k}  (\phi_{-k}^{1}c_{k}^{1}
 +\phi_{-k}^{0}c_{k}^{0}+
\phi_{-k}^{-1}c_{k}^{-1}+\phi_{-k}^{L}c_{k}^{L})\nonumber\\  &&-
{1\over 2}\sum_{k,l}\left(2ic_{-k}^{1}c_{-l}^{-1}b_{k+l}^{0}
+ic_{-k}^{1}c_{-l}^{0}b_{k+l}^{1}-ic_{-k}^{-1}c_{-l}^{0}b_{k+l}^{-1}\right.
\nonumber\\
 &&+(k+l)c_{-k}^{1}c_{-l}^{L}
b_{k+l}^{1}+(k-l)c_{-k}^{-1}c_{-l}^{L}b_{k+l}^{-1} +kc_{-k}^{0}
c_{-l}^{L}b_{k+l}^{0}\nonumber\\  &&\left.+ \half
(k-l)c_{-k}^{L}c_{-l}^{L}b_{k+l}^{L}+\mbox{h.c.}\right),
\e{53}
 where h.c. are
hermitian
conjugate terms and
\be
&&\phi^{-1}_{n}\equiv\half \sum_{k}p_{k}
\cdot p_{n-k}, \quad\phi^{0}_{n}\equiv\frac{1}{4}
\sum_{k}(x_{k}
\cdot p_{n-k} + p_{n-k}\cdot x_{k}),\nn\\ &&\phi^{1}_{n}\equiv\half
\sum_{k}x_{k} \cdot x_{n-k}, \quad\phi^{L}_{n}\equiv-i\half
\sum_{k}k(x_{k}\cdot p_{n-k} + p_{n-k}\cdot x_{k}),
\e{57}
which are the Fourier modes of the hermitian operator constraints
corresponding to
\r{51}. The BRST charge \r{53} is formally nilpotent and a consistent
BRST quantization is possible
if there exists a BRST invariant vacuum state on which we have a nilpotent  BRST
operator. In order to investigate the existence of such a vacuum state
we need the extended constraint
 operators defined by
\be
\tilde{\phi}^{-1}_{n}\equiv[{\cal{Q}},
b_n^{-1}]&=&\phi^{-1}_{n}+ \sum_{k}
\left(2ic^{1}_{k}b^{0}_{n-k}+ic^{0}_{k}b^{-1}_{n-k} -
(n+k)c^{L}_{k}b^{-1}_{n-k}\right)\nonumber
\\ \tilde{\phi}^{0}_{n}\equiv[{\cal{Q}}, b_n^{0}]&=&\phi^{0}_{n}+\half
\sum_{k}\left(ic^{1}_{k}b^{1}_{n-k}-ib^{1}_{n-k}c^{1}_{k}
-2nc^{L}_{k}b^{0}_{n-k}-\right.\nonumber \\
&&\left.-ic^{-1}_{k}b^{-1}_{n-k}+ib^{-1}_{n-k}c^{-1}_{k}\right) \nonumber \\
\tilde{\phi}^{1}_{n}\equiv[{\cal{Q}}, b_n^{1}]&=&\phi^{1}_{n}-
 \sum_{k}\left(2ic^{-1}_{k}b^{0}_{n-k}
+ic^{0}_{k}b^{1}_{n-k}+(n-k)c^{L}_{k}b^{1}_{n-k}\right)\nonumber \\
\tilde{\phi}^{L}_{n}\equiv[{\cal{Q}}, b_n^{L}]&=
&\phi^{L}_{n}- \half\sum_{k}\left(
(k+n)(c^{-1}_{k}b^{-1}_{n-k}-b^{-1}_{n-k}c^{-1}_{k}
+c^{L}_{k}b^{L}_{n-k}-b^{L}_{n-k}c^{L}_{k})+\right.\nonumber\\
&&\left.+(k-n)(c^{1}_{k}b^{1}_{n-k}-
b^{1}_{n-k}c^{1}_{k})+k(c^{0}_{k}b^{0}_{n-k}-b^{0}_{n-k}c^{0}_{k})]\right).
 \e{58}
These operators are shown to satisfy the following  commutator algebra
\be
&&[\tilde{\phi}^{1}_{m},\tilde{\phi}^{-1}_{n}]=2i\tilde{\phi}^{0}_{m+n},\qquad
 [\tilde{\phi}^{L}_{m},\tilde{\phi}^{L}_{n}]=(m-n)\tilde{\phi}^{L}_{m+n},
 \qquad
[\tilde{\phi}^{0}_{m},\tilde{\phi}^{L}_{n}]= m\tilde{\phi}^{0}_{m+n},\nonumber\\
 &&[\tilde{\phi}^{1}_{m},\tilde{\phi}^{0}_{n}]=
i\tilde{\phi}^{1}_{m+n},\qquad \qquad
 [\tilde{\phi}^{-1}_{m},\tilde{\phi}^{0}_{n}]= -i\tilde{\phi}^{-1}_{m+n},
\nonumber\\
&&[\tilde{\phi}^{1}_{m},\tilde{\phi}^{L}_{n}]=(m+n)\tilde{\phi}^{1}_{m+n},
 \qquad
[\tilde{\phi}^{-1}_{m},\tilde{\phi}^{L}_{n}]=(m-n)\tilde{\phi}^{-1}_{m+n}.
\e{581}
These commutators are non-anomalous as required by the formal nilpotence of $Q$.

Following
\cite{GLS} we consider now  a vacuum state which satisfies the conditions
\be
&&p^M_n|0\hb=b_n^{-1}|0\hb=c_n^1|0\hb=0,\quad \forall n.
\e{59} These conditions  are consistent with a BRST invariant vacuum state
since \be
[{{Q}}, p^M_n
]|0\hb=\tilde{\phi}^{-1}_{n}|0\hb=[Q, c_n^1]|0\hb=0,\quad
\forall n, \e{591}   are
satisfied due to \r{59}. Now these conditions do not specify a unique
vacuum. We need further conditions for that. We may ghost fix the vacuum by the
conditions
\be
&&b_n^0|0\hb=b_n^L|0\hb=0,\quad \forall n,
\e{514}
Their consistency conditions are
\be
&&\tilde{\phi}^{0}_{n}|0\hb=   \tilde{\phi}^{L}_{n}|0\hb= 0,\quad \forall n,
 \e{510}
and are satisfied for $n\neq0$ due to \r{59}. By means of the regularization
\r{127} we find furthermore that \be
&&\tilde{\phi}^{L}_{0}|0\hb=0
\e{511}
for any $s$ due to the symmetry properties we had in \r{47}. Using
\r{1292} we have by a direct calculation
\be
&&\tilde{\phi}^{0}_{0}|0\hb={i\over 2}(d-2)\al|0\hb
\e{512} in the limit
$s\,\ra\,0$. Thus, $\tilde{\phi}^{0}_{0}$ has an imaginary continuous
spectrum. The above vacuum state is therefore
only BRST invariant for $d=2$. (There
might exist BRST invariant states built from the
above vacuum state. However, we have
not found any.)

If we define a new charge like in the bosonic string by
\be
&&Q'=Q-{i\over 2}(d-2)\al c^0_0,
\e{513}
then the above vacuum state is BRST invariant under the new charge $Q'$ for any
dimension
$d$. However,
$Q'$ is then neither hermitian nor nilpotent for $d\neq 2$.

In fact, $Q$ is nilpotent on $|0\hb$
for any dimension $d$. This may be checked as in the previous models,
 by computing the algebra
of the extended contraints acting on the vacuum state. The most non-trivial
of these equations read
\be
[\tilde{\phi}^{-1}_{m},\tilde{\phi}^{1}_{-m}](s)|0\hb={2-d\over 2}
\sum_{k} f^{(\al)}(|k|,s)f^{(\al)}(|k+m|,s)|0\hb.
\e{5191}
Here
\be
&& \sum_{k}\f(|k|,s)\f(|k+m|,s)=\zeta(s,\alpha+\half|m|,
-{m^2\over4})+\nn\\
&&+\zeta(s,\alpha-\half|m|,
-{m^2\over4})+g(m,s),
\e{520}
where we have introduced the zeta function \cite{Eliz}
\be
&&\zeta(s,a,b)\equiv\sum_{k}{1\over[(k+a)^2+b]^s}
\e{521}
and $g(m,s)$ involves only finite sums and $g(m,0)=-1$. Now since (we are
indebted to Per Salomonson for this simple argument)
\be
&&\left.{d\zeta\over
db}\right|_{s=0}=\left.s\sum_{k}{1\over[(k+a)^2+b]^{s+1}}\right|_{s=0}=0,
\e{522}
we have
\be
&&\zeta(0,a,b)=\zeta(0,a,0)=\zeta(0|a)=\half-a,
\e{523}
which implies
\be
&&\left.\sum_{k}\f(|k|,s)\f(|k+m|,s)\right|_{s=0}=-2\alpha.
\e{524}
Hence
\be
[\tilde{\phi}^{-1}_{m},\tilde{\phi}^{1}_{-m}]|0\hb=(d-2)\al|0\hb,\e{5241}
which is consistent with \r{581} and \r{512}. Commutators of other
constraint operators may similarly be shown to consistently act on the
vacuum state in accordance with eq \r{581}. Thus the nilpotency of the BRST
operator holds as a true operator equation in the
chosen state space. For $d\neq 2$ the considered
vacuum state is not BRST invariant.
Furthermore, we have not been able to find any BRST invariant state in the state
space. If there does not exist BRST invariant
states the theory is non-trivial only
in $d=2$.

Now instead of the conditions \r{514} we may also ghost fix the vacuum
state by the
conditions
\be
&&c_n^0|0\hb=b^L_n|0\hb=0,\quad \forall n.
\e{525}
In this case all consistency conditions are satisfied, which means that
this vacuum
is BRST invariant under the original formally nilpotent BRST charge \r{53}
for any
dimension $d$. Hence, we have found two vacua: one which is BRST invariant
only for
$d=2$ and another which is BRST invariant for all dimensions. However,
since we work
in the minimal sector we must make sure that there exists a dual vacuum state
$|\bar{0}\hb$. It is straight-forward to show that $|\bar{0}\hb$ exists for any
dimensions in the first case above but only for $d=2$ in the second case.
Thus, both
solutions yield equivalent results and we have found a BRST invariant
vacuum only
for $d=2$
in agreement with the result of
\cite{GLS}. That we have two solutions requires a selection condition. (What
we have here
is a noncanonical situation in the language of \cite{Princ}.)

\setcounter{equation}{0}
\section{Alternative quantizations of the bosonic conformal string}

We will here consider two alternative set of
state spaces to the one treated in the
previous section.
\subsection{A consistent solution at $d=4$}
First we consider a state space
with a vacuum state defined through
the conditions

\be
&&p^\mu_n|0\hb=p^+_n|0\hb=x^+_n|0\hb=b_n^{-1}|0\hb=
c_n^1|0\hb=b_n^0|0\hb=b_n^L|0\hb=0,
\quad \forall n.
\e{660}
It is easily checked that this vacuum state satisfies
\be
&&\tilde{\phi}^{-1}_{n}|0\hb=   \tilde{\phi}^{L}_{n}|0\hb= 0,\quad \forall n.
 \e{661}
Furthermore,
\be
&&\tilde{\phi}^{0}_{0}|0\hb={i\over 2}(d-4)\al|0\hb
\e{662} in the limit
$s\,\ra\,0$. By the same reasoning as in the previous section this leads to the
conclusion that the above vacuum state is only
BRST invariant for $d=4$. We must also
check the closure of the extended constraints
when acting on the vacuum state. In
precisely the same way that lead to eq. \r{5241} one finds in this case
\be
[\tilde{\phi}^{-1}_{m},\tilde{\phi}^{1}_{-m}]|0\hb=(d-4)\al|0\hb,\e{663}
which is consistent with \r{581} and \r{662}. Other commutators of constraint
operators may similarly be shown to consistently
act on the vacuum state. Concluding,
we have a non-anomalous theory for any $d$ and the
vacuum state is BRST invariant for
$d=4$. We have not been able to find BRST invariant
states for other $d$. The vacuum
dual to $|0\hb$ is defined by
\be
&&x^\mu_n|\bar{0}\hb=x^-_n|\bar{0}\hb=p^-_n|
\bar{0}\hb=c_n^{-1}|\bar{0}\hb=b_n^1|\bar{
0}\hb=c_n^0|\bar{0}\hb=c_n^L|\bar{0}\hb=0,
\quad \forall n.
\e{6621}
It is easily
shown that this vacuum is BRST invariant for any $d$.

The state space that we have defined here
treats the $x^\pm$ coordinates differently than the
$x^\mu$ coordinates. Thus we have
lost manifest $d+2$ dimensional $SO(d,2)$ covariance. However, as the $x^\mu$
coordinates are regarded as the physical ones (in the sense of the gauge fixing
conditions \r{52}), we still have manifest Lorentz covariance in this physical
subspace. (If we  do not insist on manifest Lorentz covariance, we may treat
$x^0$ and one of the  space coordinates in a similar fashion as $x^\pm$ and
get a
BRST invariant vacuum in
$d=6$.)

\subsection{An inconsistent solution}
We now consider another choice of state space. In analogy with the alternative
treatment of the tensionless string in
section 5 we
may for the conformal string try a vacuum state satisfying
\be
&&p_m^M|0\hb=x_m^M|0\hb=0,\quad m>0.
\e{701}
This alternative quantization was investigated in \cite{Is}. The consistency
conditions to \r{701} requires
\be
&&c_m^{-1}|0\hb=c_m^{0}|0\hb=c_m^{1}|0\hb=c_m^{L}|0\hb=0,\quad m>0.
\e{702}
The conditions \r{701} and \r{702} allow for a ghost
fixing of
the form
\be
&&c_0^1|0\hb=0,\quad b_m^1|0\hb=0,\quad m>0,
\e{703}
\be
&&b_m^{-1}|0\hb=b_m^{0}|0\hb=b_m^{L}|0\hb=0,\quad m\geq0.
\e{704}
The consistency conditions for \r{703} are automatically satisfied as well
as those
of \r{704} for $m>0$. However,
\be
&&\tilde{\phi}^{-1}_{0}|0\hb=0
\e{705}
 and
\be
&&\tilde{\phi}^{0}_{0}|0\hb=0
\e{706}
yields further conditions on the vacuum state. They may be satisfied. (Eq.\r{706}
fixes the conformal dimension of the vacuum state.)
The problem is the last one
which yields
\be
&&\tilde{\phi}^{L}_{0}|0\hb=(d-2)\left(\half\al^2-{1\over12}\right)|0\hb.
\e{707}
Checking the commutation relations we find
\be
&&\lim_{s\ra0}[\tilde{\phi}^{L}_{m},
\tilde{\phi}^{L}_{-m}](s)=2m\tilde{\phi}^{L}_{0}+{1\over6}(d-26)m^3-m(d-2)\al^2,
\e{708}
where $\tilde{\phi}^{L}_{0}$ on the right-hand side is the normal ordered
operator in
\r{707}. Notice that \r{708} coincide almost exactly with \r{411}
although
we here have $d+2$ coordinates and 4 ghost fields. It looks now as if the
situation is
the same as for the tensionless string in section 5, \ie it looks as if we
may follow
the arguments after \r{411} leading to a consistent BRST quantization in $d=26$.
However, for the conformal string there is an additional nontrivial
commutator for
which we find
\be
&&\lim_{s\ra0}[\tilde{\phi}^{1}_{m},
\tilde{\phi}^{-1}_{-m}](s)|0\hb=\left(2i\tilde{\phi}^{0}_{0}-
{1\over2}(d-6)m\right)|0\hb,
\e{709}
where $\tilde{\phi}^{0}_{0}$ on the right-hand side is the normal ordered
operator
that follows from
\r{58} through our regularization, \ie
\be
&&\tilde{\phi}^{0}_{0}={1\over4}(x_0p_0+p_0x_0)+\half\sum_{k=1}^\infty(p_{-k
}\cdot
x_k+x_{-k}\cdot
p_k)+i(c_0^1b_0^1-c_0^{-1}b_0^{-1})+\nn\\
&&+i\sum_{k=1}^\infty\left(b_{-k}^{-1}\cdot
c_k^{-1}-c_{-k}^{-1}\cdot b_k^{-1} -b_{-k}^1\cdot
c_k^1+c_{-k}^1\cdot b_k^1\right).
\e{710}
The relation \r{709} says that $Q$ can only be nilpotent in $d=6$ which
contradicts
the above result that required $d=26$. It follows that there is no
consistent BRST
quantization of the conformal string on  the vacuum considered here. This result
agrees with \cite{Is}.


\setcounter{equation}{0}
\section{Alternative state space for the bosonic string?}
In view of the treatments in sections 4 and 5 one may wonder if one may not
have a
corresponding vacuum state also in the ordinary bosonic string case.
Below we demonstrate that this is not the case. The classical constraints
\r{301} may
also be written as
 \be
&&P^{\mu}(\sigma)P_{\mu}(\sigma) + T^{2} {X^{\mu}}^{\prime}(\sigma
)X_{\mu}^{\prime}(\sigma) =0,\quad P^{\mu}(\sigma)X^{\prime}_{\mu}(\sigma)=0.
\e{601}
The Fourier modes of the corresponding hermitian constraint operators are
\be
\Phi^{-1}_{n}&\equiv &\half
\sum_{k}p_{k}\cdot p_{n-k}-2T^{2}
\sum_{k}k (n-k)x_k\cdot x_{n-k},\nonumber\\ \phi^{L}_{n}&\equiv
&-i\half \sum_{k}k(x_{k}\cdot p_{n-k} + p_{n-k}\cdot x_{k}).
\e{602}
The hermitian and formally nilpotent BRST charge \r{305} may be rewritten as
\be {{Q}}&=& \sum_{k} (
\Phi_{-k}^{-1}c_{k}^{-1}+\phi_{-k}^{L}c_{k}^{L})-\nn\\
&&-\frac{1}{2}\sum_{k,l}(k-l)\left(c_{-k}^{-1}c_{-l}^{L}b_{k+l}^{-1}+
\half c_{-k}^{L}c_{-l}^{L}b_{k+l}^{L}+
4T^{2}c^{-1}_{-k}c^{-1}_{-l}b^{L}_{k+l}+ \right.\nonumber\\
&&\left. +b_{k+l}^{-1}c_{-k}^{-1}c_{-l}^{L}+\half
b_{k+l}^{L}c_{-k}^{L}c_{-l}^{L}\right ).
\e{603}
(Notice that the ghost variables $c^L_k$, $b^L_k$ here are not the same as in
\r{305}.) The
extended constraints are here given by
\be
\tilde{\phi}^{L}_{n}&\equiv &[Q,
b^{L}_n]=\phi^{L}_{n}-\frac{1}{2}\sum_{k} (n+k)(c_{k}^{-1}
b_{n-k}^{-1}-b_{n-k}^{-1}c_{k}^{-1}+c_{k}^{L}
b_{n-k}^{L}-b_{n-k}^{L}c_{k}^{L}),\nonumber\\
\tilde{\Phi}^{-1}_{n}& \equiv &
[Q,b^{-1}_n]=\Phi^{-1}_{n}-\sum_{k} (n+k)(c_{k}^L
b_{n-k}^{-1}+4T^2c_{k}^{-1} b_{n-k}^{L}),
\e{604}
A direct calculation of the commutator $\left[
\tilde{\phi}^{L}_{m},\tilde{\phi}^{L}_{-m}\right]$ gives
\be
&&\left[
\tilde{\phi}^{L}_{m},\tilde{\phi}^{L}_{-m}\right](s) =\nonumber\\ &&
m\sum_{k}\left (-ikx_k\cdot
p_{-k}-kc^{L}_kb^{L}_{-k}-kc^{-1}_kb^{-1}_{-k} \right ) \left (\f(|m+k|,s)+
\f(|m-k|,s) \right )+\nonumber\\ && +\sum_{k}\left
(-ik^2 x_k\cdot
p_{-k}+(2m^2-k^2)(c^{L}_kb^{L}_{-k}+c^{-1}_kb^{-1}_{-k})\right )\left
(\f(|m+k|,s)- \f(|m-k|,s) \right ).\nonumber
\ee
The $s\ra0$ limit yields
\be &&\left[
\tilde{\phi}^{L}_{m},\tilde{\phi}^{L}_{-m}\right] =  2m\tilde{\phi}^{L}_{0},
\e{605}
which is consistent with the fact that the BRST charge \r{603} is nilpotent.
One may note here that the
second line in the last relation  goes to zero as
$s$ goes to zero if the operator
\be
\Psi_{k,m}\equiv-ik^2 x_k\cdot
p_{-k}+(2m^2-k^2)(c^{L}_kb^{L}_{-k}+c^{-1}_kb^{-1}_{-k})
\e{606}
 is well-defined.
This is true for a state space where the vacuum
has the form of the tensionless string vacuum defined by the condition
\be p_n^{\mu}|0\rangle =0,\qquad \forall n.
\e{607}
  However this vacuum state is not appropriate for the
 tensile bosonic string. Consistency conditions of the form
\be [{{Q}},p_{n}^{\mu}]|0\rangle =0,\quad \forall n \ee would
require the ghost part of such a state to satisfy the relations
\be c^{-1}_{n}|0\rangle
=0,\quad \forall n
\e{608}
for which the consistency conditions,
\be
&&[Q, c^{-1}_{n}]|0\hb=0,\quad \forall n
\e{609}
are  satisfied. In fact, it is possible to ghost fix this vacuum state in a BRST
invariant way using the original BRST charge which then also is nilpotent
for any
dimension
$d$. However, there exists no dual vacuum state, $|\bar{0}\hb$, to this
solution.
Note that $|\bar{0}\hb$ must satisfy
\be
&&x_n^{\mu}|\bar{0}\hb=b_n^{-1}|\bar{0}\hb=0
\e{610}
for which the consistency condition
\be
&&[Q, b_n^{-1}]|\bar{0}\hb=0
\e{611}
cannot be satisfied.

\setcounter{equation}{0}
\section{Alternatives with negative norm states.}
In the alternative treatments in section 5 and 7  we considered a vacuum
state of the type
\be
&&p_m^{\mu}|0\hb=x_m^{\mu}|0\hb=0, \quad m>0.
\e{801}
It lead to a ``consistent" BRST quantization for the bosonic
tensionless string in $d=26$. In fact, even the ordinary bosonic string has
a vacuum
state satisfying \r{801}. To see this one may notice that \r{801} is
equivalent to (We let $|0\hb$ be an eigenstate of $p^{\mu}_0$, \ie we set
$|0\hb=|p\hb$.)
\be
&&\al_m^{\mu}|p\hb=0,\quad \tilde{\al}_{-m}^{\mu}|p\hb=0,\quad m>0.
\e{802}
due to the relations \r{303}. Now \r{802} is different from the standard
vacuum \r{309}. The consistency conditions of the latter condition requires
\be
&&c_{-m}^K|p\hb=0,\quad m>0
\e{803}
which means that the vacuum may be ghost fixed by
\be
&&b_{-m}^K|p\hb=0,\quad m\geq0
\e{804}
together with
\be
&&c_{m}^L|p\hb=0,\quad m>0;\quad b_{m}^L|p\hb=0,\quad m\geq0.
\e{805}
Calculating commutators for finite $s$ and then taking the limit $s\ra0$ we find
\be
&&\left [\tilde{K}_{m},\tilde{K}_{-m}\right
]=2m\tilde{K}_{0}-{1\over12}(d-26)m^3+\half m(d-2)\al^2,
\e{806}
where
\be
&&\tilde{K}_{0}=\half \alpha^{2}_0
+\sum_{k=1}^{\infty}(\tilde{\alpha}_{k}\cdot\tilde{\alpha}_{-k}
-kb^{K}_{k}c^{K}_{-k}-kc^{K}_{k}b^{K}_{-k})-\half(d-2)\left(\half\al^2-
\frac{1}{12}\right).
\e{807}
We may then define a new BRST charge by
\be
&&Q'=Q-{d-2\over4}\al^2(c_0^L-c_0^K),
\e{808}
which then is nilpotent for $d=26$.  In $d=26$ the new extended constraints are
\be
&&\tilde{L}'_{0}=\half \alpha^{2}_0
+\sum_{k=1}^{\infty}(\alpha_{-k}\cdot\alpha_{k}
+kb^{L}_{-k}c^{L}_{k}+kc^{L}_{-k}b^{L}_{k})-1,\nn\\
&&\tilde{K}'_{0}=\half \alpha^{2}_0
+\sum_{k=1}^{\infty}(\tilde{\alpha}_{k}\cdot\tilde{\alpha}_{-k}
-kb^{K}_{k}c^{K}_{-k}-kc^{K}_{k}b^{K}_{-k})+1,
\e{809}
However, $\tilde{L}'_{0}|p\hb=0$ requires $p^2=8T$ while $\tilde{K}'_{0}|p\hb=0$
requires
$p^2=-8T$. It follows that neither of these vacua are BRST invariant, or in
other
words we have no BRST invariant vacuum satisfying
\r{801}. Still there are BRST invariant states and they are massless with
spin 0,1,
and 2, and spinless with $p^2=\pm 8T$. Note, however, that since
$\tilde{\al}_{m}^{\mu}$, $m>0$, act as creation operators the space components
$\tilde{\al}_{m}^{i}$ yield negative normed states. Thus, the BRST
invariant state
space have negatively normed states. The same feature was also found for the
alternative treatment of the tensionless string in section 5. (This alternative
treatment has only massless states
\cite{GRR2}.)
It is clear that the existence of indefinite metric states in the BRST invariant
sector makes both these models inconsistent.

\bigskip
\begin{flushleft}
{\bf Acknowledgments}
P.S. wish to thank  R. vonUnge and U. Lindstr\"om for
useful comments and suggestions. The work of P.S. was supported by the European
Commission TMR Programme under the grant ERBFMBI-CT97-2823.
\end{flushleft}



\begin{thebibliography}{Simple}

\bibitem{Ste}
M. Kato and K. Ogawa, \ {\sl Nucl. Phys.}\ {\bf B212}, 443 (1983)\\ S.
Hwang, \ {\sl  Phys. Rev.}\ {\bf D28}, 2614 (1983)

\bibitem{FMS}D. Friedan, E. Martinec and S. Shenker,  \ {\sl Nucl. Phys.}\
{\bf B271},
93 (1986)


\bibitem{Is}
 J. Isberg, {\it ``Tensionless Strings with Manifest Space-Time Conformal
Invariance''} {USITP-92-10}  (1992)

\bibitem{GLS}
  H. Gustafsson, U. Lindstr\"om, P. Saltsidis, B. Sundborg and R.v. Unge,
{\em Nucl. Phys.\/} {\bf B440} (1995) 495.


\bibitem{Princ}S. Hwang
and R. Marnelius, \  {\sl Nucl. Phys.}\ {\bf B315}, 638 (1989);
{\em ibid.}\ {\bf B320},\ 476\ (1989)


\bibitem{Aux}
S. Hwang, \ {\sl  Nucl. Phys.}\ {\bf B322}, 107 (1989)\\
R. Marnelius, \ {\sl Nucl. Phys.}\ {\bf B372}, 218 (1992);\ {\bf B384}, 318
(1992)

\bibitem{BAF} I. A. Batalin and E. S. Fradkin, {\sl Ann. Inst. Henri
Poincar\'e}\
{\bf
49}, 145-214 (1988)


\bibitem{Eliz}
E. Elizalde et al, \ {\em Zeta Regularization Techniques with Applications},\\
 \ {\sl World Scientific Pub. Co..}\ (1994)

\bibitem{tension}
 A. Schild, {\it Phys. Rev.} {\bf D16} (1977) 1722.\\
 A. Karlhede and U. Lindstr\"om, {\it
Class. Quant. Grav.} {\bf 3} (1986) L73.\\
 F. Lizzi, B. Rai, G. Sparano and A. Srivastava,
{\it Phys.Lett.} {\bf 182B} (1986) 326.\\
R. Amorim and J. Barcelos-Neto, {\it Z.Phys.}{\bf
C38} (1988) 643.\\
U. Lindstr\"om, B. Sundborg and G. Theodoridis,
{\it Phys. Lett.}  {\bf 253B} (1991) 319.\\
 U. Lindstr\"om, B. Sundborg and G. Theodoridis,
{\it Phys. Lett.} {\bf 258B} (1991) 331.



\bibitem{dtwo}
 F. Lizzi, B. Rai, G. Sparano and A. Srivastava,
{\it Phys.Lett.} {\bf 182B} (1986) 326.\\
 J. Isberg, U. Lindstr\"om, B. Sundborg and G.
Theodoridis, {\it Nucl. Phys.}, {\bf B411} (1994) 122.

\bibitem{GRR}J. Gamboa, C. Ramirez, and M. Ruiz-Atlaba,
 \ {\sl  Phys. Lett.}\ {\bf 225B}, 335 (1989)

\bibitem{Gauge}R. Marnelius,
 \ {\sl Nucl. Phys.}\ {\bf B412}, 817 (1994),\\ I. A. Batalin and R.
Marnelius, \ {\sl Nucl. Phys.}\ {\bf B442}, 669 (1995)

\bibitem{Pan} P. Saltsidis, {\em Nucl. Phys.} {\bf B446} (1995) 286

\bibitem{Pan2} P. Saltsidis, {\em Phys. Lett.} {\bf 401B} (1997) 21

\bibitem{WW}E. T. Whittaker and G. N. Watson, \ {\em Modern Analysis}. \ (fourth
edition),\\
 \ {\sl Cambridge University Press.}\ (1952)

\bibitem{GRR2}J. Gamboa, C. Ramirez, and M. Ruiz-Atlaba,
\ {\sl Nucl. Phys.}\ {\bf B338}, 143
(1990)

\bibitem{MN}R. Marnelius, \  {\sl  Phys. Rev.}\ {\bf D20}, 2091 (1979),\\
R. Marnelius and B. Nilsson, \  {\sl  Phys. Rev.}\ {\bf D22}, 830 (1979)


\end{thebibliography}
\end{document}